\documentclass[reprint,aps,amsmath,amssymb,]{revtex4-2} 
\usepackage{graphicx}
\usepackage{dcolumn}
\usepackage{bm}

\begin{document}

\title{Two-qubit atomic gates: Spatio-temporal control of Rydberg interaction}

\author{Ignacio R. Sola}
\email{corresponding author: isolarei@ucm.es} 
\affiliation{Departamento de Quimica Fisica, Universidad Complutense, 28040 Madrid, Spain}
\author{Vladimir S. Malinovsky}
\affiliation{DEVCOM Army Research Laboratory, 2800 Powder Mill Road, Adelphi, Maryland 20783, USA}
\author{Jaewook Ahn}
\affiliation{Department of Physics, Korea Advanced Institute of Science and Technology (KAIST), Daejeon 34141, Republic of Korea}
\author{Seokmin Shin}
 \affiliation{School of Chemistry, Seoul National University, 08826 Seoul, Republic of Korea}
\author{Bo Y. Chang}
 \email{corresponding author: boyoung@snu.ac.kr}
 \affiliation{School of Chemistry, Seoul National University, 08826 Seoul, Republic of Korea}
 \affiliation{Research Institute of Basic Sciences, Seoul National University, 08826 Seoul, Republic of Korea}

\begin{abstract}
By controlling the temporal and spatial features of light, we propose a novel protocol to prepare two-qubit entangling gates on atoms trapped at close distance, which could potentially speed up the operation of the gate from the sub-micro to the nanosecond scale.
The protocol is robust to variations in the pulse areas and the position of the atoms, by virtue of the coherent properties of a dark state, which is used to drive the population through Rydberg states.
From the time-domain perspective, the protocol generalizes the one proposed by Jaksch and coworkers \ [Jaksch \textit{et al., Phys. Rev. Lett.}, 2000, \textbf{85}, 2208], with three pulses that operate symmetrically in time, but with different pulse areas. 
From the spatial-domain perspective, it uses structured light.
We analyze the map of the gate fidelity, 
which forms rotated and distorted lattices 
in the solution space.
Finally, we study the effect of an additional qubit to the gate performance and propose generalizations that operate with multi-pulse sequences.
\end{abstract}

\maketitle

\section{Introduction}

Due to their excellent optical addressability, \cite{Kuzmich_PRL2022,Kuzmich_Science2012,Adams_JPB2019,Adams_PRL2010,Ohmori_NPhoto2022}  
rich many-body physics,\cite{Saffman_PRL2007,Lukin_PRL2010,Lukin_PRL2001,Saffman_PRA2015,Zhan_PRL2017,Lukin_PRL2018,Saffman_NPhys2009}
and long decoherence times,\cite{Saffman_RMP2010}
neutral atoms excited in Rydberg states 
have been used as vectors for different quantum technologies,
including entanglement preparation, \cite{Grangier_PRL2010,Saffman_PRA2010,Walker_AAMOP2012,Grangier_NPhys2009,Picken_QCT2018,Saffman_Nature2022,Malinovskaya_OL2017,Malinovskaya_SR2021}
creation of photonic entanglement,~\cite{Thomas_Nature2022} quantum simulators,~\cite{Weimer_NPhys2010,Lukin_Nature2017,Lukin_Nature2019,Ahn_PRL2018} 
and quantum computation.\cite{Jaksch_PRL2000,Saffman_RMP2010,Saffman_QIP2011,Saffman_Nature2022,Grangier_PRL2012,Saffman_JPB2011,Zoller_PRL2009,Shi_PRApp2018,Molmer_PRX2020,Li_PRApp2021}

A key step involves trapping the atoms at low temperature in magneto-optical traps (MOT). 
Homogeneous magnetic fields working on the MOT split the 
degeneracy of the hyperfine ground states of the atoms,
allowing to encode and address separately the qubit states 
by optical fields through intermediate 
states or using microwave fields.\cite{Chu_PRL1985,Chu_PRL1987,Chu_OptLett1992}
Currently, it is possible to create splittings $\Delta$ of the order of $10$ GHz\cite{Grangier_PRL2010}
that in principle, allow to drive the population from the $|0\rangle$
state independently of the $|1\rangle$ state in the nanosecond regime.

On the other hand,
due to the strong dipole-dipole interaction of Rydberg states, \cite{Adams_book2018,gallagher_1994}
the energy of a double excitation of Rydberg states
$|rr\rangle$ is larger than twice the energy of each atom in a Rydberg
state. For weak enough pulses (or close enough atoms)
this extra energy is also larger than the Rabi frequency driving the transition, and the $|rr\rangle$ state
cannot be populated. This defines a maximum distance, called the Rydberg blockade radius, 
$R_{\cal B}$, within which the well-known C-PHASE gate protocol was proposed by Jaksch et al. \cite{Jaksch_PRL2000}.
The three-pulse sequence protocol consists of a $\pi$-pulse acting on the first qubit, followed by a $2\pi$-pulse acting on the second qubit, and a $\pi$-pulse again acting on the first qubit.
The pulse frequencies are tuned to excite the chosen Rydberg state, $|r\rangle$
from the qubit state $|0\rangle$ (alternatively, from the $|1\rangle$ state) so the other
qubit state is decoupled.
Then if the system is initially in the $|00\rangle$ state, the first pulse moves
the amplitude to $i|r0\rangle$, the second does nothing and the third moves
the amplitude to $-|00\rangle$.
If the system starts in $|01\rangle$, the first pulse acts as before, driving
the amplitude to $i|r1\rangle$, the second does nothing and the third
induces the transition to $-|01\rangle$.
When the system is in $|10\rangle$ the first pulse does nothing, the
second drives the amplitude to $-|10\rangle$ and the third does nothing.
Finally, the lasers cannot induce any transition from the $|11\rangle$ state.
In the following, we will refer to this set of operations as the Jaksch protocol (JP).

There have been several proposals to extend the JP mechanism  
using more robust adiabatic excitation schemes\cite{Goerz_PRA2014,Bergmann_RMP1998,Bergmann_ARPC2001}
adding alternative processes to the dipole blockade through dark states,\cite{Molmer_PRA2017}
or addressing multi-qubit gate generalizations.\cite{Molmer_PRX2020,Li_PRApp2021,Saffman_PRL2019,Lukin_PRL2019}
Some of these ideas extend well known optical control adiabatic strategies 
\cite{Malinovsky_PRA2014,Malinovsky_PRA2004,Malinovsky_PRL2004,Malivsky_PRL2006}
for dynamics with target states conditioned on the initial state.
\cite{Kosloff_PRA2003,Goerz_NJP2014,Caneva_PRA2011,Goerz_JPB2011}
One disadvantage of these schemes is the need to work with long
pulses, in the microsecond regime.
In the JP scheme, this is needed to operate with independent qubits,
forcing interatomic distances of the order of $\sim 5 \mu$m, with Rydberg-Rydberg
interactions, $d_{\cal B}$, of a few MHz.
The time-scale gap in $\Delta^{-1}$ and $d^{-1}_{\cal B}$ offers an opportunity
to speed the gates typically by two orders of magnitude,
by using denser arrays of atoms, therefore boosting the
dipole blockade such that $d_{\cal B} \sim \Delta$.
Although the physics of Rydberg states is rich and several unwanted physical processes 
may be involved in working in denser media, \cite{Saffman_PRA2016,Grangier_NPhys2009,Comparat_JOSAB2010,Tong_PRL2004,Saffman_PRA2008,Pohl_PRL2009,Sadeghpour_PRL2010,Sadeghpour_NComm2018}
it is important to estimate if the JP
can operate under these conditions or if one can design other robust protocols for the
gates with non-independent qubits. 

One way to extend the JP to more compact arrangements of atoms
is to implement control not only in time (encoded in the pulse sequence), but also in space
(encoded as geometrical parameters) taking into account 
the strength of the light-matter interaction
at the exact qubit locations.
In the simplest arrangements, the spatial control might just involve specific focusing 
of each laser light at different points of the lattice formed by the atoms, not just necessarily 
at the site that the atom occupies, 
using various TEM modes of light.\cite{Murty_AppOpt1964}

Here we examine the potential advantages of using so-called structured light for the entangling gate implementation.
Recent advances in the control of spatial properties of light have been used to
create topological electromagnetic effects, 
as well as almost arbitrary complex geometrical patterns
that can be complemented with pulse shaping techniques.\cite{Forbes_NPhoto2021,Rubinsztein_Dunlop_JOpt2016,Ohmori_NPhoto2022} 
For instance, a superposition of TEM$_{00}$ and TEM$_{01}$ or TEM$_{10}$
can be used to create nodes and phase relations between the peak values
of the fields at nearby locations.
In addition to the space-dependent intensities, such {\em hybrid modes}
(see the sketch in Fig.\ref{proposal}) 
may have different phases
at different positions of the atoms.
The idea is to use a symmetrical protocol where the first pulse and its copy,
focusing on the first qubit, are applied before and after a second pulse
acts mostly on the second qubit. We 
prove that a particular relation in
the spatial features of the pulses confers special robustness to the gate
implementation, giving raise to the {\em Symmetric Orthogonal Protocol} or SOP
that is proposed in this work.
By incorporating the basic features of spatio-temporally controlled pulses as parameters in the Hamiltonian, we design 
simple models to estimate the fidelities 
for two- and three-qubit systems, implementing
the C-PHASE type gate.
The complexity of the system increases with the number of parameters, 
but what can be seen as a drawback, may 
be seen as an opportunity to find novel ways to
control the system 
using  optimization techniques.

Our study is a first demonstration
of quantum control application\cite{Rice_2000,Shapiro_2011,Chang_IJQC2016} to design quantum gates 
addressing both the
spatial and temporal features of the laser fields.
Using simplified models at zero temperature and without noise, 
we 
show that the spatio-temporal control of the fields acting
locally on
each qubit can be used to achieve robust and efficient implementations of fast entangling gates. 
In fact, this theoretical setup offers many more practical implementations to control
gates and prepare entanglement, which will be addressed in the future. 
The speed of the designed gates justifies some our assumptions, like neglecting decoherence, dephasing, and most 
mechanisms that might lead to fidelity losses. 
However, more detailed studies considering
the effect of other states in the system, the presence of Stark shifts, noise in the pulses, and the motion of the atoms, 
are required to assess
the practical implementation of the SOP.

\section{Gate performance for non-separated qubits: Analysis}

The computational basis of a two-qubit in our system is composed of
$|00\rangle$, $|01\rangle$, $|10\rangle$ and $|11\rangle$ states.
Together with the Rydberg ancillary states: $|r0\rangle$, $|0r\rangle$,
$|r1\rangle$, $|1r\rangle$, form the basis where we will follow
the evolution of the system, as we assume that the $|rr\rangle$ state
cannot be accessed by strong Rydberg blockade.
We assume that the distance between atoms is shorter than the waist of the laser beams.
As a first approximation to model the local effect of the field 
on each of the qubits, we define {\em geometrical factors}, $a_k$ and $b_k$, 
so the spatially and temporally dependent interaction of the laser $k$ at qubit
$\alpha$ ($\alpha = a,b$) is determined by the Rabi frequencies 
$\Omega_k(\vec{r}_\alpha,t) = \alpha_k\mu_{0r} E_k(t)/\hbar$.
The geometrical factors can be partially incorporated in the Franck-Condon factor 
$\mu_{0r}$ so we can assume, without loss of generality, that $a_k$ and $b_k$ 
are normalized to unity ($\sqrt{a_k^2 + b_k^2} = 1$). 
Using hybrid modes of light (structured light) 
one can control $a_k$ and $b_k$ in a wide range of values,
including negative factors.

\begin{figure}
\includegraphics[width=8.0cm]{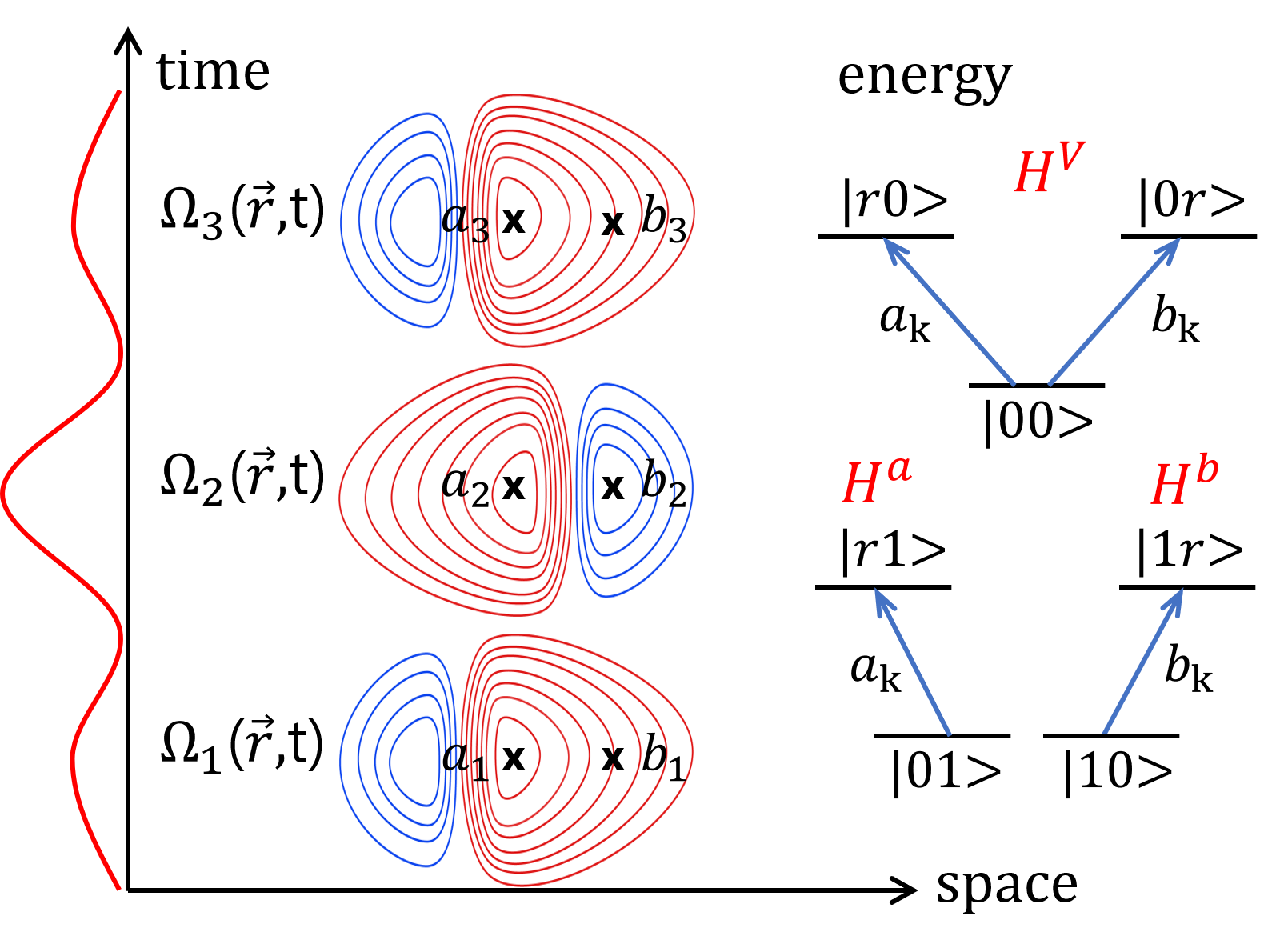}
\caption{Proposed implementation of the SOP for two-qubit gates. 
The qubits are located at the $x$ positions. 
They are subject to the pulses, $\Omega_1(\vec{r},t)$,
$\Omega_2(\vec{r},t)$, $\Omega_3(\vec{r},t)$ whose spatial profile is shown in
the horizontal ``{\em space}'' axis, with local amplitudes at the qubits designated
by the geometrical factors $a_k$ and $b_k$. 
Different colors represent a $\pi$ phase change in the amplitude of the field (red - positive; blue - negative).
The sequence of operations that
governs the temporal evolution of the state vectors depends on the pulse sequence,
shown in the vertical ``{\em time}'' axis. 
Conditional on the initial state of the qubit, $\lbrace |00\rangle, |01\rangle,
|10\rangle \rbrace$, the energy diagram of the subsystem where the dynamics
takes place is shown under ``energy''. The $|11\rangle$, not shown, 
is completely decoupled.}
\label{proposal}
\end{figure}

Using a pulse sequence of non-overlapping pulses $\Omega_k(\vec{r},t)$, 
in resonance between
the $|0\rangle$ state of the qubit and the chosen Rydberg state $|r\rangle$, the
Hamiltonian is block-diagonal, 
${\sf H}_k^V \oplus {\sf H}_k^a \oplus {\sf H}_k^b \oplus {\sf H}^d$,
where 
${\sf H}_k^V = -\frac{1}{2} \Omega_k(t) \left(
a|00\rangle \langle r0| + b |00\rangle
\langle 0r| + \mathrm{h.c.} \right)$
is the Hamiltonian of a $3$-level system in $V$ configuration,
acting  in the subspace of $\lbrace |00\rangle, |r0\rangle, |0r\rangle \rbrace$
states, 
${\sf H}_k^a = -\frac{a}{2}\Omega_k(t) \left(  
|01\rangle\langle r1| + \mathrm{h.c.} \right)$
and 
${\sf H}_k^b = -\frac{b}{2} \Omega_k(t) \left(  
|10\rangle\langle 1r| + \mathrm{h.c.} \right)$ 
are two-level Hamiltonians acting 
in the subspace of $\lbrace |01\rangle, |r1\rangle \rbrace$ and
$\lbrace |10\rangle, |1r\rangle \rbrace$ respectively.
Finally, ${\sf H}^d = 0$
is a zero Hamiltonian acting on the state $|11\rangle$, decoupled from any field.
The energy diagrams of the subsystems are shown in Fig\ref{proposal}.

The time-evolution operator of any of
these Hamiltonians can be solved analytically through their 
time-independent dressed states, that have zero non-adiabatic couplings.\cite{shore_2011,Malinovsky_2011,Sola_AAMO2018}
For reference, we include here the analytical form of the time-evolution operators 
at the end of the pulses.
For ${\sf H}_k^V$,
\begin{equation}
U^{V}_{k} = 
\left( \begin{array}{ccc}
\cos \theta_k & i  a_{k} \sin \theta_k & i b_{k} \sin \theta_k  \\
i a_{k} \sin \theta_k & a_{k}^2 \cos \theta_k + b_k^2 & 
a_{k}b_{k} \left[ \cos \theta_k - 1 \right]  \\
i b_{k} \sin \theta_k & a_{k}b_{k} \left[ \cos \theta_k - 1 \right]  & 
b_{k}^2 \cos \theta_k + a_k^2 \end{array} \right)  \label{UV}
\end{equation}
where the mixing angle
$$\theta_k = \frac{1}{2} \int_{-\infty}^{\infty}\Omega_{k}(t) dt = \frac{1}{2} A_k$$
is half the pulse area.
Interestingly, ${\sf H}_k^V$ supports a dark state, 
$|\Phi^0_k\rangle = -b_k |r0\rangle + a_k|0r\rangle$.
This is a dressed state of zero energy, 
which is uncoupled to the ground state, $|00\rangle$, that is, $|\Phi^0_k\rangle$
cannot decay nor be excited by the field. On the other hand, the time-evolution
operator that affects the initial states $|01\rangle$ and $|10\rangle$, is
\begin{equation}
U^{\alpha}_{k} =  \left( \begin{array}{cc}
\cos \vartheta_k & i \sin \vartheta_k \\
i \sin \vartheta_k & \cos \vartheta_k  \end{array} \right)
\label{U2LS}
\end{equation}
where $\vartheta_k = \alpha_k \theta_k$ $(\alpha = a, b)$.

\subsection*{Effect on $|00\rangle$}

Let us predict the effect of the JP on the partially distinguishable qubits.
Excitation from $|00\rangle$ with a $\pi$-pulse prepares 
the entangled state $|\Psi_1\rangle = i a_1 |r0\rangle +  i b_1|0r\rangle$. 
Because there is population in both qubits and $\Omega_2(\vec{r},t)$
acts on both, the second pulse can 
induce transitions from $|\Psi_1\rangle$,                   
breaking  the mechanism of the JP. 
If we want $|\Psi_1\rangle$ to remain undisturbed, we need 
$|\Psi_1\rangle$ to be the dark state of ${\sf H}_2^V$, $|\Phi_2^0\rangle$, 
for which $a_1 = -b_2$ and $b_1 = a_2$. 
Then the third pulse can be another copy of the first pulse, transforming 
$|\Psi_1\rangle$ back into $-|00\rangle$.

To generalize the result we define {\em structural vectors}, which are 
normalized two-component vectors  formed by the geometrical factors, 
$\vec{e}_k = \left( a_k, b_k \right)$ that characterize the spatial properties of the protocol.
Success in the gate performance (starting in $|00\rangle$) implies that the first diagonal 
element of the full propagator, $U_{11}^{V} = \left( U^V_3 U^V_2 U^V_1 \right)_{11} = -1$.
Working out the matrix multiplication we obtain
\begin{equation}
\begin{array}{ll}
U_{11}^{V} & = c_3 c_2 c_1 - (\vec{e}_2\vec{e}_1) c_3 s_2 s_1 
-(\vec{e}_3\vec{e}_2) s_3 s_2 c_1    \\  & %
-(\vec{e}_3\vec{e}_2)(\vec{e}_2\vec{e}_1) s_3 c_2 s_1 
-\left[ \vec{e}_3\vec{e}_1 - (\vec{e}_3\vec{e}_2)(\vec{e}_2\vec{e}_1) \right] s_3 s_1
\end{array}  \label{U11V}
\end{equation}
where we have used the compact notation $c_k = \cos\theta_k$, $s_k = \sin\theta_k$.  
This formula is valid in general
for a $3$-pulse sequence on arrays of $N$ qubits, where $\vec{e}_k$ becomes a $N$-dimensional
vector.
The condition that $|\Psi_1\rangle = |\Phi_2^0\rangle$ (the dark state of ${\sf H}^V_2$) 
is that $\vec{e}_2$ is orthogonal to both $\vec{e}_1$ and $\vec{e}_3$. 
In symmetric protocols, with $\vec{e}_3 = \vec{e}_1$ and $\theta_1 = \theta_3$, one obtains
the equation that defines the behavior under the SOP,
\begin{equation}
U_{11}^{V} = \cos^2\theta_1 \cos\theta_2  - \sin^2\theta_1 \ .
\label{U11Vs}
\end{equation}
For $2$-qubit systems, the normalization and orthogonality of the structural vectors
implies that there is only one free parameter, $b \equiv b_1 = -a_2$, that 
measures the overlap of the field on both qubits and as such, indirectly
measures the proximity of the atoms in the trap.

If $\theta_1 = \pi/2$, as in the JP, $U_{11}^{V} = -1$ regardless of $\theta_2$
and for any value of $b$.
In addition, Eq.(\ref{U11Vs}) guarantees a remarkable robustness to variations
in the pulse areas.
Making $\delta\theta^\prime_1 = \pi/2 + \delta_1 = \theta^\prime_3$, 
and $\theta^\prime_2 = \pi + \delta_2$, one can easily obtain
\begin{equation}
U_{11}^{\prime^{V}} = \delta_1^2 \left( -1 + \frac{\delta_2}{2}\right)
- \left( 1 - \frac{\delta_1^2}{2} \right)^2 = -1 + (2\delta_2^2-\delta_1^2) \frac{\delta_1^2}{4} \ .
\label{U11robust}
\end{equation}
Any error in the pulse areas will only add a quartic error in $U_{11}^V$  
($\sim\!\delta^4$).

\subsection*{Effect on $|01\rangle$ and $|10\rangle$}

To evaluate the gate performance when the two-qubit system is initiated in $|01\rangle$ 
or $|10\rangle$ we use Eq.(\ref{U2LS}). 
In a symmetrical sequence, the final state will be ($\alpha = a, b$)
\begin{equation}
U_{11}^{\alpha} = \left( U^\alpha_3 U^\alpha_2 U^\alpha_1 \right)_{11} =
\cos \left( 2\alpha_1 \theta_1 + \alpha_2 \theta_2 \right) \ .
\label{U11alpha}
\end{equation}
In Jaksch protocol, ($\theta_1 = \pi/2$, $\theta_2 = \pi$),
$U_{11}^{\alpha} = \cos \left( [\alpha_1 + \alpha_2] \pi \right)  = -1$
for independent qubits, as only one component, $\alpha_1$ or $\alpha_2$, exist.
But for orthogonal geometrical factors ($a_1 = b_2 \equiv a, b_1 = -a_2 \equiv b$),
\begin{equation}
U_{11}^{a} +  U_{11}^{b} = 2\cos\left( a\pi\right) \cos\left( b\pi\right)
\approx -2 +b^2\pi^2
\end{equation}
inducing quadratic deviations in $b$ that lower the fidelity of the gate.

A compromise must be made in the choice of the pulse areas.
Because $U_{11}^V$ does not depend on the geometrical factors (nor on $A_2$), 
the pulse parameters 
should be adjusted mainly due to their effect on $U_{11}^a$ and $U_{11}^b$.
which require ``rotating'' the pulse areas.
From Eq.(\ref{U11alpha}), for any pulse area in the SOP, we obtain
\begin{eqnarray}
U_{11}^a = \cos\left( a A_1 - \frac{1}{2} b A_2 \right) = \cos\left( \frac{1}{2} \left[
a A_\mathrm{odd} - b A_\mathrm{even} \right]\right) \label{cosrot1} \\ 
U_{11}^b = \cos\left( b A_1 + \frac{1}{2} a A_2 \right) = \cos\left( \frac{1}{2} \left[
b A_\mathrm{odd} + a A_\mathrm{even} \right] \right)
\label{cosrot2}
\end{eqnarray}
where we use even and odd pulse areas.
We can write Eqs.(\ref{cosrot1},\ref{cosrot2}) in terms of new mixed pulse areas
$A^\prime_\mathrm{odd}$, $A^\prime_\mathrm{even}$, where
\begin{equation}
 \left( \begin{array}{c} A^\prime_\mathrm{odd} \\ A^\prime_\mathrm{even} \end{array} \right) = \left( \begin{array}{cc}
a  & -b \\ b & a \end{array} \right)  \left( \begin{array}{c} A_\mathrm{odd} \\ 
A_\mathrm{even} \end{array} \right)  \label{RA}
\end{equation}
such that $A^\prime_\mathrm{odd} = 2\pi 
(1 + 2m)$, $A^\prime_\mathrm{odd} = 2\pi 
(1 + 2n)$, as in the JP, but for the new mixed pulse areas. Although $U_{11}^V$ will affect the overall fidelity, we observe that the effect of using orthogonal geometrical vectors is to rotate
clockwise
by an angle $\beta = \arctan ( b / a)$,
the pulse areas that will give
maximal but not perfect fidelities.

\section{Gate performance for non-separated qubits: Numerics}

\begin{figure}[h]
\includegraphics[width=7cm]{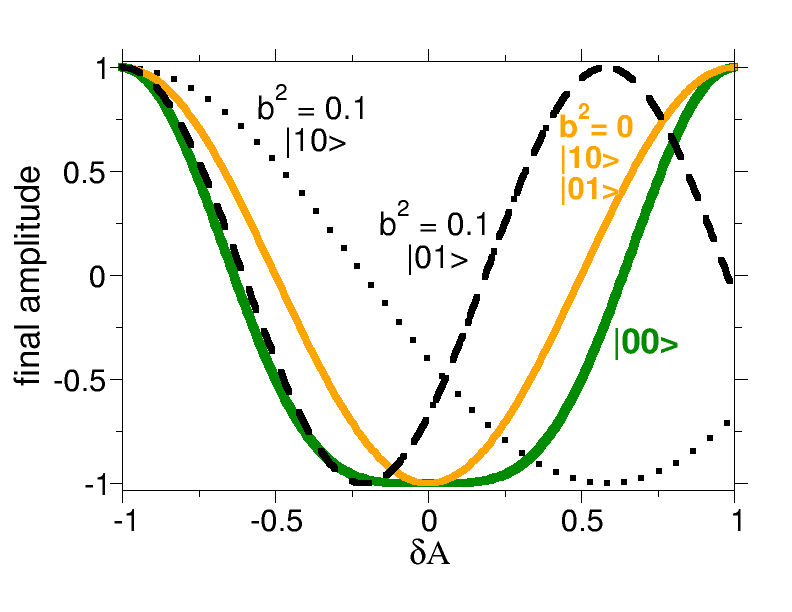}
\caption{Dependence of the final amplitude of the starting 
state  ($|00\rangle$, $|01\rangle$ and $|10\rangle$ on deviations from $\pi$ in the pulse areas of odd pulses ($\delta A_1 = \delta A_3 = \delta A$), and of $2\pi$ in the second pulse ($\delta A_2 = 2\delta A$), for different geometrical factors.} 
\label{UvsA}
\end{figure}

As discussed in the previous section, the orthogonality of the geometrical factors
assures that population passage in the $V$ subsystem comprising
the $|00\rangle$, $|r0\rangle$ and $|0r\rangle$ states, goes through a
dark state. It is so robust that errors in the pulse areas only imply quartic
deviations in $U^{V}_{11}$ regardless of the value of $b$, that is, of the proximity of the qubits.  
This is shown In Fig.\ref{UvsA} (solid line), where $U_{11}$ are computed
for the different subsystems as a function of variations in pulses areas from the
JP, assuming $\delta\!A = \delta\!A_1  = \delta\!A_2 /2$.
On the other hand, $U^{\alpha}_{11}$ increases quadratically around the
minima (dashed black lines in Fig.\ref{UvsA}) and become disaligned for 
$|01\rangle$ and $|10\rangle$ as $b$ increases.
Thus, the desired value of $U_{11}^\alpha =-1$ may fall outside
of the flat band that characterizes the behavior of $U_{11}^V$ around its minima,
disrupting the gate's efficiency.

\begin{figure}
\hspace*{-0.5cm}
\includegraphics[width=9.5cm]{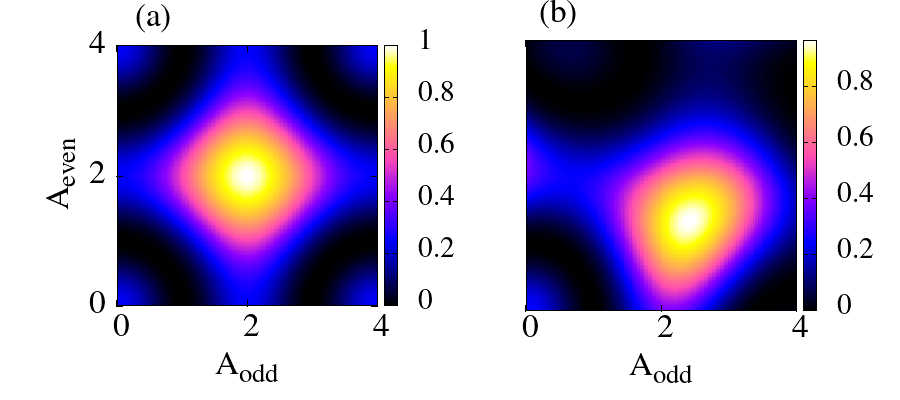}
\caption{Gate fidelity 
of the (a) JP and (b) SOP for $b^2=0.1$, as a function of the pulse areas
(in units of $\pi$) for the solutions with minimal pulse area.}
\label{FvsA2detail}
\end{figure}

To optimize the gate one needs to change the pulse areas in specific directions.
The smallest optimal pulse areas that maximize the fidelity are shown in
Fig.\ref{FvsA2detail}, for $b^2 = 0.1$.
To avoid a distortion in the map of solutions (henceforth {\em fidelity map})
we represent the map as a function on the area acting on each qubit,
$A_\mathrm{odd} = A_1 + A_3$ and $A_\mathrm{even} = A_2$, rather than $A_1$ and $A_2$.
We obtain a maximum fidelity of $0.96$
displaced to larger areas in $A_\mathrm{odd}$ and smaller areas in $A_\mathrm{even}$.
The overall pulse area, $A_T = |A_\mathrm{odd}| + |A_\mathrm{even}| = 3.7\pi$,
is however smaller than in the JP ($4\pi$).
On the other  hand, the overall robustness of the gate, measured as the area occupied by the maximum in the space of solutions, is practically the same. 

\begin{figure}
\includegraphics[width=8.0cm]{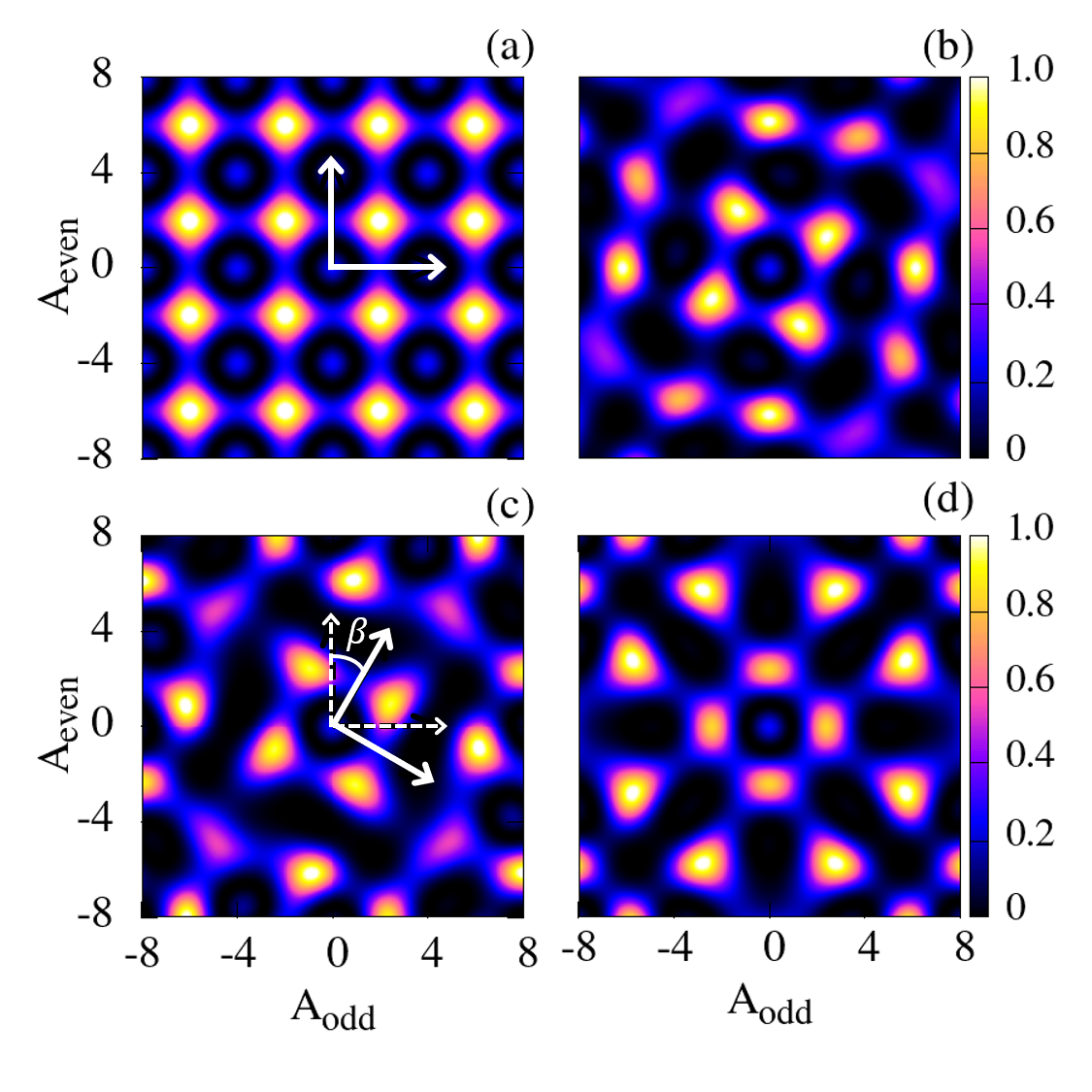}
\caption{Fidelity map as a function of the pulse areas (in units of $\pi$) for  different geometrical factors: (a) $b^2 = 0$, (b) $b^2 = 0.1$, (c) $b^2 = 0.2$, (d) $b^2 = 0.5$.}
\label{FvsA2}
\end{figure}

But this is only the minimal pulse area implementation. In the JP, other solutions are possible modulo area $2\pi$ in the areas of the first and third pulses, $A_1$ and $A_3$, and modulo $4\pi$ in $A_2$. Hence the family of protocols with symmetric pulses ($A_3 = A_1$) satisfying $A_\mathrm{odd} = 2\pi (1 + 2m)$ and $A_\mathrm{even} = 2\pi (1+2n)$, where $m, n$ are integers, give perfect fidelity in the absence of noise or perturbations.
While in the JP all the different protocols give the same fidelity, this is not the
case for the SOP.
Exploring solutions for larger pulse areas, better fidelities ($F \ge 0.98$) are found,
as shown in Fig.\ref{FvsA2}.

The fidelity map for the JP 
is a regular lattice with spacings $\Delta A_\mathrm{odd} = \Delta A_\mathrm{even}
= 4\pi$.
In the SOP, the lattice is rotated with respect to the JP, with
a rotation angle of $\beta = \arctan(b/a)$, in agreement with Eq.(\ref{RA}). 
There are some distortions as $b^2$ increases, regaining
a perfect, but different, symmetrical pattern as one reaches $b^2 = 0.5$ ($\beta = \pi/4$). 
All maps for different $b$ have approximately the same number of maxima,
separated a minimum distance of $4\pi$,
so the density of solutions is conserved.
However, except for the $b=0$ case, not all the fidelities at the maxima
reach unity.
Typically, larger areas are needed to find better solutions, {\it e.g.}
$A_\mathrm{odd} = -6.1\pi$, $A_\mathrm{even} = 0.9\pi$ for a total
area of $A_T = 7\pi$ and a peak fidelity of $F = 0.99$ for $b^2 = 0.2$.
On the other hand, the minimal pulse area at which the first maxima appears
decreases with $b$, and for $b^2 = 0.5$ one observes 
$F = 0.8$ using $\Omega_2(\vec{r},t)$ only, or $\Omega_1(\vec{r},t)$ 
and $\Omega_3(\vec{r},t)$, for a total area of only $2.42 \pi$.

\begin{figure}
\centering
\includegraphics[width=8.0cm]{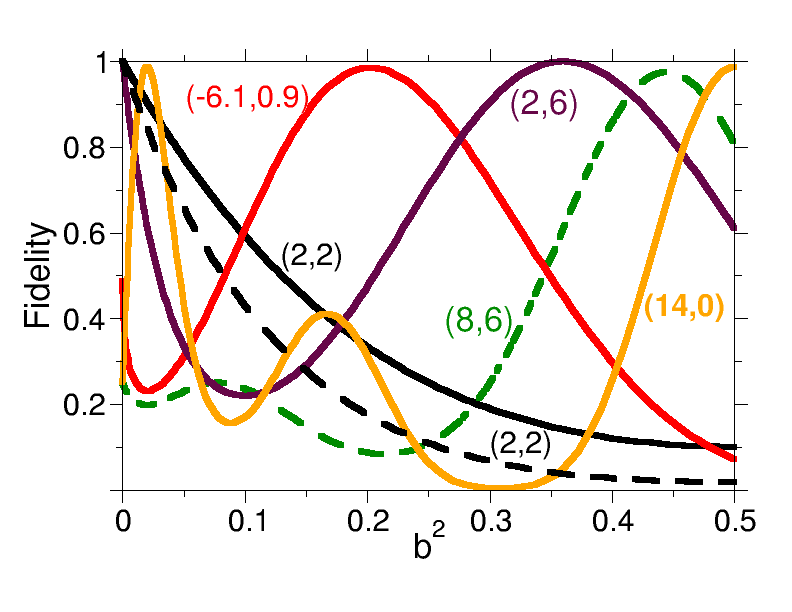}
\caption{Dependence of the fidelity of the gate on the geometrical factor $b$
for different values of the pulse areas. Using the minimal pulse areas of the
JP, we show how the fidelity decays more slowly along the SOP
protocol conditions (black line) than when
the atoms become closer without imposing the orthogonality (black dashed line).}
\label{Fvsb}
\end{figure}

Instead of plotting how the fidelity map varies
as a function of the pulse areas
for fixed geometrical factors, we can fix the areas and vary $b$.
This is done in Fig.\ref{Fvsb}.
Only a few possible choices for the pulse areas are shown.
The black line shows how the fidelity falls with the JP parameters, 
$A_1 = \pi$, $A_2 = 2\pi$, as $b$ increases. In terms of $A_\mathrm{odd}$
and $A_\mathrm{even}$ we classify this protocol with a couple of numbers
$(A_\mathrm{odd},A_\mathrm{even}) = (2,2)$ in units of $\pi$.
Other implementations of the JP scheme without minimal pulse areas, as $(2,6)$,
decay faster than the $(2,2)$ because of the larger accumulated pulse area, but surprisingly recover
and work perfectly at different values of $b$.
Protocols that do not belong to the JP, as $(8,6)$, where $A_\mathrm{odd}$
is not of the form $A_\mathrm{odd} = 2\pi (1 + 2m)$, fail at $b=0$ but also provide high fidelities
at certain values of $b$.

Although natural multiples of $\pi$ for the pulse areas usually work relatively well, the
ratios between the areas do not need to be natural numbers. The number of solutions
is dense. 
For instance, taking into account the rotation of the optimal pulse areas depending on
$b$, we show how the fidelity changes for the protocol $(-6.1, 0.9 )$, which maximizes
the fidelity at $b^2 = 0.2$ [see Fig.\ref{FvsA2}(c)].
More surprisingly, one also finds solutions that do not
require the three-pulse strategy, like $(14,0)$ in which the second pulse does
not participate, but nevertheless one achieves
high fidelities at low and large $b$.
These protocols are related to approximate solutions of
Diophantine equations, where the gate mechanism does not rely on the dark
state. They will be explored elsewhere.

The curves in Fig.\ref{Fvsb} show that the fidelities decay following a quadratic behavior
when the geometrical factors depart from the optimal values, as expected.
However, the decay is often slower at larger $b$. 
(And the effect is even more noticeable if $F$ is plotted against $b$, instead of $b^2$.) 
Interestingly, the SOP guarantees a slower  decay in the presence of the other
qubit. Compare the black line with the black dashed line.
In the black dashed line we start in the JP conditions and show what happens
when the atoms approach (the qubits are no longer independent) but
we do not use structured light. Then $\vec{e}_2 = (b, a)$ instead of $(-b, a)$.
In solid black line we use the SOP. As long as the displacement of the atoms preserves the orthogonality of the structural vectors, the decay in the fidelity of the SOP is clearly slower.

\section{Three-qubit systems}

\begin{figure}
\hspace*{-0.5cm}
\includegraphics[width=9.5cm]{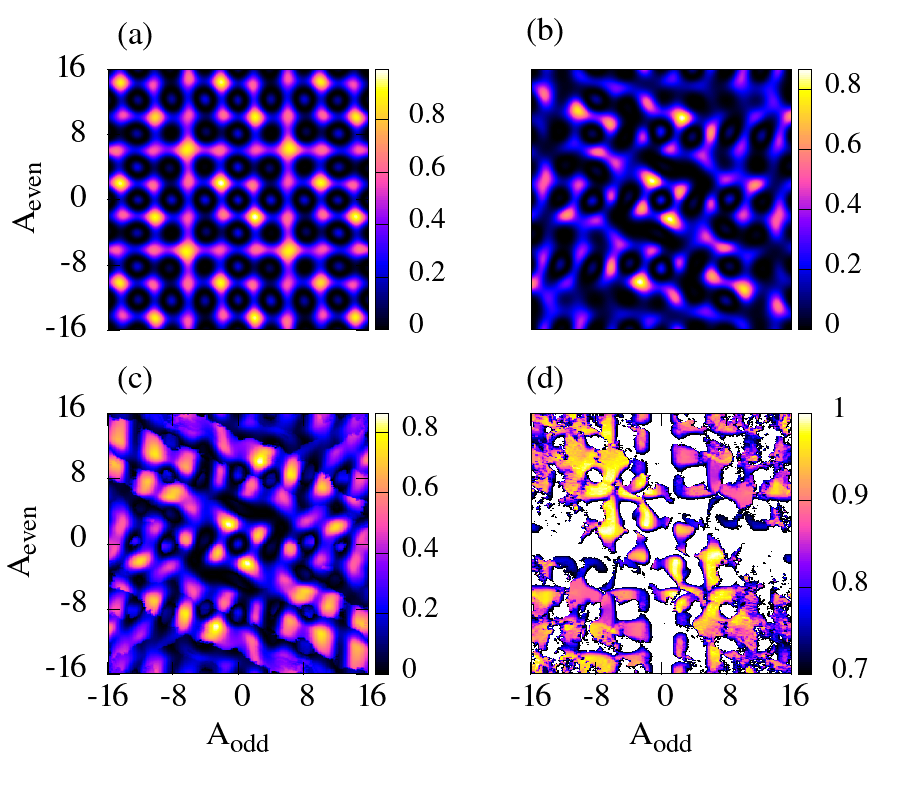}
\caption{Fidelity map of the ${\cal P}_{ab}^-$ gate as a function 
of the pulse areas (in units of $\pi$) in a 3-qubit system.
(a) JP scheme with $c_k^2 = 0.1$ for all pulses. (b) SOP with 
$b^2=0.1$ and $c_k^2=0.1$. (c) SOP 
with optimized $c_k$ imposing symmetric and orthogonal conditions
in qubits $a$ and $b$.
(d) Fidelity performing at optimized $a_k$ and $b_k$ for 
$c^2_k = 0.1$, imposing symmetric conditions.}
\label{F3qvsA}
\end{figure}

The protocol proposed in this work implies the use of denser arrays of trapped atoms.
It is then important to analyze its robustness to the presence of other
qubits not involved in the two-qubit gate. 
In this section, we check the efficiency of the SOP to prepare the gate
${\cal P}^-_{ab}$, which is the C-PHASE gate acting on qubits $a$  and $b$
in the presence of qubit $c$. This gate must operate exactly as with two
qubits, regardless of the state of the third qubit, hence
the diagonal of the ${\cal P}^-_{ab}$ matrix has the signature
$\mathrm{diag}\lbrace -1, -1, -1, -1, -1, -1, 1, 1 \rbrace$ when
the basis is ordered as $\lbrace |000\rangle, |010\rangle, |100\rangle,
|001\rangle, |101\rangle, |011\rangle, |110\rangle, |111\rangle \rbrace$.
The theoretical treatment follows closely the analysis in Section.2 and
the main equations (\ref{U11Vs}), (\ref{U11alpha}) remain valid by adding a
third component, $c_k$, which gives the geometrical factor on qubit $c$,
to the structure vector $\vec{e}_k$.

For independent qubits, as in the JP, the result is the same
as for two qubits, see Fig.\ref{FvsA2}(a).
However, if the third qubit is close to the other two 
($c_k^2 = 0.1$ for all pulses), even if the second and first qubits are sufficiently apart 
($b^2=0$), the fidelity
already decreases as shown in Fig.\ref{F3qvsA}(a).
The lattice of solutions looks like in the two-qubit system, 
but some local maxima can be rather smaller than one,
although high fidelity solutions still exist.

On the other hand, if one starts with nearby qubits $a$ and $b$ and
implements the SOP, is the presence of qubit $c$
more disrupting? The answer is in Fig.\ref{F3qvsA}(b), where we assumed
that $c_k^2 = 0.1$ for all pulses, and we applied the SOP to the remaining
qubits forcing orthogonality and symmetric conditions.

As in the $2$-qubit case, the fidelity map 
is rotated by 
approximately the same angle as before, with  a slight shift due to $c_k$.
But the highest fidelities are now clearly smaller than one
($F_\mathrm{max} \sim 0.85$), making the protocol less useful.

What if the $c_k$ parameters are optimized?
Can one control the position of qubit $c$ (or the spatial profile
of the laser located in this qubit) such that the fidelity increases,
for fixed values of the other geometrical parameters?
Fig.\ref{F3qvsA}(c) shows that the third qubit cannot be used to
improve the fidelity of the ${\cal P}_{ab}^-$ gate. 
Here we project at every value of
$(A_\mathrm{odd}, A_\mathrm{even})$ the best fidelity obtained by optimizing $c_k$, using symmetric
pulses [$\Omega_3(t) = \Omega_1(t)$] and
forcing orthogonal conditions over the subspace of the first two qubits, 
as in SOP. The parameters are found by using a simplex optimization with linear constraints.\cite{Nelder_CJ1965}
The optimization does not change drastically the fidelity map, 
which shows the same angle of rotation,
but the picture becomes blurry. By controlling
$c_k$ (with $c_k^2 \ge 0.1$) the protocols that appear as local
maxima become more robust (the peaks are like plateaux)
and the fidelities at low maxima increase,
but the highest fidelities are still $F \sim 0.85$.

Perfect fidelities can be obtained for most pulse areas in the SOP if
the geometrical factors of {\em all} qubits involved are controlled
(forcing a minimal value of $\alpha_k^2 \ge 0.1$ and keeping $c_k = 0.1$ fixed). 
This is shown in
Fig.\ref{F3qvsA}(d), where the minimum fidelity chosen is $0.7$
in the map (close to
the maximum fidelity in Fig.\ref{F3qvsA}(b)).
The patterns of solutions display a regular, non-rotated lattice, 
similar to that in Fig.\ref{F3qvsA}(a), but with
plateaux rather than maxima, that show high fidelities everywhere except for
small pulse areas. 
Even at minimal pulse area of $A_T = 4\pi$ one can find protocols with $F \ge 0.99$.
Clearly, there are many protocols that implement efficient and robust 
$2$-qubit gates in systems of $3$ not fully distinguishable qubits, but
they require finer control. In particular, in optimizing all the geometric
parameters we enforce the symmetry ($\vec{e}_3 = \vec{e}_1$)
but not the orthogonalization, so the working protocols are not necessarily of the SOP type.

\section{Multipulse sequences}

\begin{figure}
\hspace*{-0.5cm}
\includegraphics[width=9.5cm]{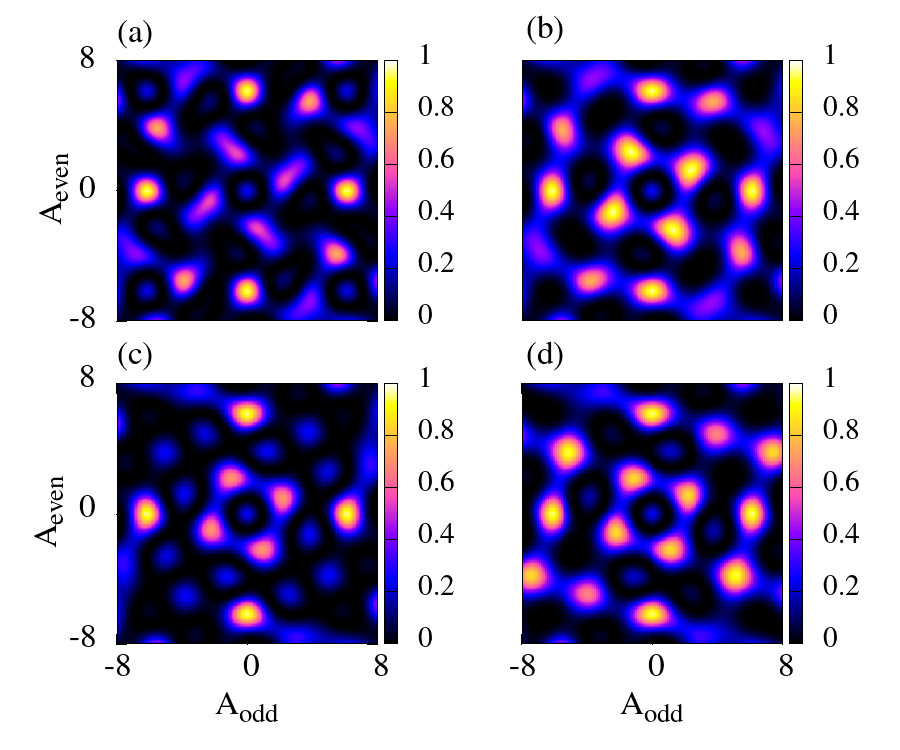}
\caption{Fidelity map for $M$-pulse sequence extensions of the SOP with
(a) M=2, (b) M = 3, (c) M = 4, (d) M = 5. We  fixed $b^2 = 0.1$ in
all cases.}
\label{FNvsA}
\end{figure}

The SOP protocol follows a set of rules that can be easily extended to
other multipulse sequences, with $M$ number of pulses. 
We here propose the {\em Extended Symmetric
Orthogonal Protocol} or ESOP, with the following features:
\begin{enumerate} 
\item
In the $M$-pulse sequence
all odd pulses are equal to each other (copies of the same
pulse) as well as all even pulses: $A_{k+2} = A_k$,
$\vec{e}_{k+2} = \vec{e}_k$. 
\item
Odd and even pulses are orthogonal to each other:  $\vec{e}_{k+1} \cdot
\vec{e}_k = 0$.
\end{enumerate}
Under these conditions it can be shown that the number of surviving terms in
$U_{11}^V$ [see Eq.(\ref{UV})] is minimal.
In particular, for 2-pulse sequences, there is only one term:
$U_{11}^V = \cos\theta_2 \cos\theta_1$.
For $3$ and more pulses, there are always terms involving products of
the sines of the areas of even or odd pulses:
\begin{align}
&\textnormal{For M=3: } U_{11}^V = \cos^2 \theta_\mathrm{odd} \cos\theta_\mathrm{even} 
- \sin^2 \theta_\mathrm{odd} \\
&\textnormal{For M=4: } U_{11}^V = \cos^2 \theta_\mathrm{odd} \cos^2\theta_\mathrm{even} 
- \sin^2 \theta_\mathrm{odd} - \sin^2 \theta_\mathrm{even} \\
&\textnormal{For M=5: } U_{11}^V = \cos^3 \theta_\mathrm{odd} \cos^2\theta_\mathrm{even} 
- 3\sin^2 \theta_\mathrm{odd} - \sin^2 \theta_\mathrm{even} 
\end{align}
where $\theta = A / 2$.
On the other hand, for the $|10\rangle$ and $|01\rangle$ states, 
the unitary evolution terms can always be written as
\begin{equation}
U_{11}^\alpha = \cos\left( \sum_k^M \alpha_k \theta_k \right)
\hspace{0.5cm} (\alpha = a, b).
\end{equation}

For $M=2$ there are no terms depending on $\sin\theta_\mathrm{odd}$, constraining too
much the possible solutions. In the JP and, to a
lesser extent in the SOP with three pulses, this was the main term that forced
the pulse areas of the {\em odd} pulses to be odd multiples of $\pi$,
leading to optimal fidelities. The operating mechanism for the
gate performance with two pulses is different.
Still, it is possible to achieve high fidelities in a smaller set of protocols. 

Fig.\ref{FNvsA}(b) is the same result shown in Fig.\ref{FvsA2}(b), repeated here
to facilitate the comparison.
Surprisingly, for $M=4$ the map is similar to that with $M=2$ in spite of
$U_{11}^V$ having more terms. The symmetry of the cosine and sine
terms on $\theta_\mathrm{odd}$ and $\theta_\mathrm{even}$ 
effectively constraints the possible solutions so the fidelity map is similar
to that with $M=2$.
The same will happen in all sequences with even number of pulses.
Solutions with odd number of pulses 
always provide richer fidelity maps, 
with higher fidelities available at more protocols.
Although solutions can be easily
generalized to any number of pulses, the highly-constraining nature of the 
ESOP schemes
makes these protocols probably unnecessary, 
as they do not improve the results of the SOP.
This might not be the case, however, when the effects of
noise are taken into account. Intensity fluctuations that are proportional
to the peak intensity of the lasers will affect more strongly those
protocols that use stronger fields, that is, with larger pulse areas.
The effect will be dominated by the pulse with larger area in the
sequence, rather than by the sum of all pulse areas. It is possible
to find optimal protocols with $5$-pulses that distribute the pulse
area among all the fields, so that the peak intensities in each field 
are smaller than in the similar protocol with $3$-pulses.
Obviously, if the parameters are unconstrained and full optimization
is performed to find the solutions, then having more pulses
and more parameters will clearly provide more protocols, as will
be shown in subsequent studies.

\section{Summary and Discussion}

We have studied the performance of a two-qubit gate in a denser array of 
trapped atoms under perfect conditions (zero temperature, no external noise)
in the regime of strong Rydberg blockade.
Extending the well-known Jaksch protocol (JP), we have proposed a novel
implementation, called the Symmetric Orthogonal Protocol (SOP), by controlling
both the pulse sequence and the spatial properties of the fields, using
structured light. 
The Hamiltonian of the light-atomic system was modeled using geometrical factors,
that measure the amplitudes of the fields at the location of the qubits,
allowing to obtain analytical formulas for the propagators.

From the time-domain perspective, the scheme generalizes the JP with
three pulses that operate symmetrically in time, but with different pulse areas.
From the spatial-domain perspective, the scheme uses hybrid modes of light. 
The geometrical factors form orthogonal vectors in the SOP,
which allow to decouple the effect of odd and even pulses in the sequence,
by using a coherent dark state that drives the population through the Rydberg states.
The SOP protocol is as robust as the JP to variations in the pulse areas and
more robust to changes in the position of the atoms along certain directions.
Implementations with maximal fidelity form a lattice of solutions in the space
of the pulse areas, which is rotated with respect to the lattice of solutions in the JP.

We have analyzed the effect of adding a third atom in the proximity of the two-qubit system. The fidelity in the SOP decays more rapidly than in the JP. 
High-fidelity solutions could not be found by just controlling the position of the 
third atom, but rather the geometrical factors at the first two-qubits must
be optimized for every choice of pulse areas.
Finally, we have proposed natural generalizations of the SOP to multipulse sequences,
showing that  sequences with odd number of pulses form richer lattices with
a denser number of solutions than sequences with even number of pulses.

The SOP shows great promise for possible implementations of fast two-qubit
gates. Working in the strong dipole blockade regime ($d_{\cal B} > \Delta$)
one can in principle accelerate the gate a factor of $200$.
However, high fidelities are achieved  in the SOP using larger pulse areas 
than in the JP, typically by a factor of $2$ to $10$.
Hence, gates with comparable or slightly worse fidelity and equal robustness 
could be in principle prepared at durations of the order of $20$ to $100$ times shorter, 
moving the scale from the microsecond to the nanosecond regime.

From the physical point of view, the SOP operates as the JP,
so one can expect a similar sensitivity to decoherence and noise.
However, because the atoms are much closer, the dipole blockade is much larger
and the gate time is much shorter, the effect of the thermal motion of the
atoms, Rydberg-Rydberg couplings or spontaneous emission, is almost negligible.
Only fluctuations in the field intensities
(hence pulse areas) as well as in the position of the atoms, leading to changes in the geometrical factors, have some impact on the fidelities.
Preliminary estimates of these effects 
show about $1$\% reduction of the peak fidelities working
at $\sim 25\mu$K temperature. However, further comprehensive studies are needed to properly quantify the effect of noise.
Equally important will be to assess the current limitations 
(time durations, spatial resolution) in using structured light
specific to the proposed design. Hybrid modes of light may find many
applications in quantum technologies, beyond the present proposal.
On the other hand,
full optimization of both the pulse areas and the geometrical factors
will probably allow refinements in the implementations, the surface
of which has only been explored in the present contribution.

\section*{Conflicts of interest}
There are no conflicts to declare.

\section*{Acknowledgements}
This research was supported by the Quantum Computing Technology Development Program (NRF-2020M3E4A1079793) and the National Creative Research Initiative Grant (NRF-
2014R1A3A2030423). IRS also thanks the BK21 program (Global Visiting Fellow) for the stay during which this project started and for support from MINECO CTQ2015-65033-P and MINECO PID2021-122796NB-I00. SS acknowledges support from the Center  for Electron Transfer funded by the Korea government(MSIT)(NRF-2021R1A5A1030054). 

\bibliography{SolaNanoscale23-arxiv} 

\end{document}